\documentclass[structabstract]{aa}
\usepackage{subfigure}                                   
\usepackage{graphicx}
\usepackage{txfonts}
\begin{document}
   \title{Intranight optical variability of radio-quiet BL Lacertae objects}


   \author{Yuan Liu
          \inst{1}
          \and
          Jin Zhang\inst{2}
          \and
          Shuang Nan Zhang\inst{1, 2}
          }

   \institute{Key Laboratory of Particle Astrophysics, Institute of High
Energy Physics, Chinese Academy of Sciences, P.O.Box 918-3, Beijing
100049, China\\
              \email{liuyuan@ihep.ac.cn; zhangsn@ihep.ac.cn}
         \and
National Astronomical Observatories, Chinese Academy of Sciences, Beijing 100012, China\\
             \email{zhang.jin@hotmail.com}}

   \date{...}


  \abstract
   {}
   {Intranight variation (or microvariation) is a common phenomenon of radio-loud BL Lac objects.
   However, it is not clear whether the recently found radio-quiet BL Lac objects have the same properties.
   The occurrence rate of intranight variation is helpful in distinguishing the mechanism of the continuum of radio-quiet BL Lac objects.}
   {We conducted a photometric monitoring of 8 radio-quiet BL Lac objects by the Xinglong 2.16m and Lijiang 2.4m telescopes.
   The differential light curves are calculated between each target and two comparison stars. To quantify the variation, the significance of variation is examined by a scaled $F$-test.}
   {No significant variation is found in the 11 sessions of light curves of 8 radio-quiet BL Lac objects (one galactic source is excluded).  The lack of microvariation in radio-quiet
   BL Lac objects is consistent with the detection rate of microvariation in normal radio-quiet AGNs, but
   much lower than for radio-loud AGNs. This result indicates that the
   continua of the radio-quiet BL Lac objects are not dominated by jets that will induce frequent microvariations.
   }
   {}


   \maketitle
%

\section{Introduction}
Active galactic nuclei (AGNs) are characterized by their broad band
continua, strong emission lines, and fast variability.
However, a handful of abnormal AGNs have recently been discovered in the SDSS data, i.e., weak line quasars and radio-quiet BL Lac objects (Diamond-Stanic et al. 2009; Plotkin et al. 2010). The
UV/optical emission lines are absent in their UV/optical
spectroscopies, though the shape and luminosity of their continua
are comparable to the normal AGNs. The fraction of these
special AGNs is small ($\sim$1/1000) in the SDSS DR 7 sample; however, it could be an important stage in the evolving sequence of AGNs (Hryniewicz et al. 2010; Liu \& Zhang 2011). In the early stage of an active cycle of AGNs, the radiative feedback can expel the gas from broad line regions and result in weak or even in the disappearance of broad emission lines (Liu \& Zhang 2011). Other models, such as of a cold accretion disk, an extremely high accretion
rate, shielding gas, or abnormal BL Lac objects, have also been proposed (Shemmer et al. 2010; Plotkin et al. 2010; Laor \& Davis 2011; Wu et al. 2012). The origin of such weak-line
AGNs is still not clear. Both the thermal
(accretion disk) or non-thermal (jet) component may explain the weak
line feature.

The optical continua of classical BL Lac objects are dominated by the
synchrotron emission from the relativistic jets, therefore high
polarization and fast variability are important
characteristics of classical BL Lac objects. Heidt
\& Nilsson (2011) found the polarization degrees of the radio-quiet BL Lac
 candidates are low. Plotkin et al. (2010) investigated the long-term
  variability of radio-quiet BL Lac objects using the  data from SDSS
 stripe 82 and find that the variation amplitude of radio-quiet BL Lac objects is
 smaller than that of radio-loud BL Lac objects. However, owing to the
 small size of their sample, this is not very conclusive.

 The
 variation in short time scale (intranight) is another
 characteristic of classical BL Lac objects. During a very short
 period, e.g. several hours,  the flux of classical BL Lac object
 can change by several tenths of a magnitude (Wagner
\& Witzel 1995; Heidt
\& Wagner 1996; Bai et
al. 1998; de Diego et al. 1998). However, it is still
 unclear whether the intranight variation is frequent in the
 radio-quiet BL Lac objets, which will be helpful in distinguishing the
 origin of their continua. Gopal-Krishna et al. (2013, hereafter GJC2013) and Chand et al. (2014, hereafter CKG2014) claim that they have detected a considerable fraction of intranight variation in a sample of weak-line AGNs (duty cycle$\sim$5\%), and this fraction can be higher if the signal-to-noise ratio of the light curve is further increased. However, as we discuss in this paper, some galactic sources may contaminate their sample.

In this paper, we report the result of our monitoring campaign  of radio-quiet BL Lac objects. Observations and data reduction are described in Section 2, and then the significance of the intranight variation is shown in Section 3. Section 4 discusses the implication of our results and presents the conclusions.

\begin{table*}[!htp]
\caption{Information on observations.} \label{table:1}
\begin{center}
\begin{tabular}{c c c c c c c c c c}     %
\hline\hline
Object (SDSS) &   RA  &   DEC &   $R$  &   Redshift    &   Date    &   Telescope\tablefootmark{a}   &   Filter  &   Duration (h)    &   N\tablefootmark{b}  \\
\hline
J081250.80+522530.8	&	123.212	&	52.425	&	17.85
	&	1.152	&	2011.2.12	&	L	&	$R$	&	4.4	&	27	\\
	&		&		&		&		&	2012.1.27	&	L	&	$R$	&	5.7	&	32	\\
J085025.60+342750.9	&	132.607	&	34.464	&	18.51
	&	1.389	&	2011.2.13	&	L	&	$R$	&	3.3	&	20	\\
J090107.64+384658.8	&	135.282	&	38.783	&	17.87
	&	unknown	&	2012.1.29	&	L	&	$R$	&	6.9	&	32	\\
J094533.99+100950.1	&	146.392	&	10.164	&	17.45
	&	1.662	&	2011.2.10	&	L	&	$V$	&	6.1	&	25	\\
	&		&		&		&		&	2012.1.28	&	L	&	$R$	&	6.0	&	34	\\
J125219.48+264053.9	&	193.081	&	26.682	&	17.51
	&	1.292	&	2011.4.23	&	X	&	$R$	&	6.3	&	38	\\
	&		&		&		&		&	2013.4.14	&	X	&	$R$	&	8.4	&	47	\\
J132809.59+545452.8	&	202.040	&	54.915	&	17.59
	&	2.096	&	2011.4.24	&	X	&	$R$	&	4.6	&	48	\\
J134601.29+585820.1	&	206.505	&	58.972	&	17.46
	&	1.667	&	2011.4.25	&	X	&	$R$	&	4.1	&	18	\\
J142943.64+385932.2	&	217.432	&	38.992	&	17.26
	&	0.930	&	2013.4.13	&	X	&	$R$	&	6.2	&	33	\\

\hline
\end{tabular}
\end{center}
\tablefoottext{a}{L$-$Lijiang 2.4m, X$-$Xinglong 2.16m}\\
\tablefoottext{b}{Number of exposures}
\end{table*}


\section{Observations and data reduction}
The amplitude of intranight variation is normally several tenths of a magnitude,  so the desired error of our observed magnitude is $\lesssim$0.05 mag with the exposure time not longer than 10 min. The corresponding magnitude threshold is $R\sim18.5$ for the 2 m class telescopes we used. Seven radio-quiet BL Lac objects were selected from Plotkin et
al. (2010), and SDSS J094533.99+100950.1 was selected from Hryniewicz et al. (2010). The above selection criteria are similar to those in GJC2013 and CKG2014. Actually, five sources in our sample are shared with GJC2013 and CKG2014 who based their selection primarily on classification by Plotkin et al. (2010) as a `high-confidence BL Lac candidate'. We additionally included in our sample some low-confidence BL Lac candidates.  The sources are classified as low-confidence only because the continuum near the emission line is hard to define, and the equivalent width of emission lines will be larger or smaller than 5 ${\AA}$ depending on the continuum assumptions, which is mainly due to the noisy spectra around some emission lines. We therefore think there should be no systematic difference between high- and low-confidence sources and will investigate the variation property of subsamples in future works.

 The observations were carried out by BFOSC (BAO Faint Object Spectrograph and Camera)  on the Xinglong (China) 2.16 m telescope and YFOSC (Yunnan Faint Object Spectrograph and Camera
) on the Lijiang (China) 2.4 m telescope. All observations
were performed in Johnson R band, except for SDSS J094533.99+100950.1
in Johnson V band. The exposure time was 300 s or 600 s depending
on the weather conditions.
 The detailed information about the sample and observations is shown in
Table 1. In total, there are 11 sessions of light curves of these eight sources.

\begin{table*}[]
\caption{Information on targets and their companion stars.}
\label{table:1}
\begin{center}
\begin{tabular}{c c c c c c}
\hline\hline
Object (SDSS)	&	Date	&	RA (J2000)	&	DEC (J2000)	&	$r$ (SDSS)	&	$g-r$ (SDSS)	\\
\hline
J081250.80+522530.8	&	2011.2.12	&	08 12 50.80	&	+52 25 30.8	&	18.05	&	0.3	\\
Star 1	&		&	08 12 51.29	&	+52 26 46.4	&	17.28	&	1.4	\\
Star 2	&		&	08 12 49.52	&	+52 26 26.2	&	17.89	&	1.3	\\
J081250.80+522530.8	&	2012.1.27	&	08 12 50.80	&	+52 25 30.8	&	18.05	&	0.3	\\
Star 1	&		&	08 12 51.29	&	+52 26 46.4	&	17.28	&	1.4	\\
Star 2	&		&	08 12 49.52	&	+52 26 26.2	&	17.89	&	1.3	\\
J085025.60+342750.9	&	2011.2.13	&	08 50 25.60	&	+34 27 50.9	&	18.66	&	0.4	\\
Star 1	&		&	08 50 26.96	&	+34 26 35.9	&	19.22	&	1.0	\\
Star 2	&		&	08 50 17.77	&	+34 26 50.5	&	18.53	&	0.7	\\
J090107.64+384658.8	&	2012.1.29	&	09 01 07.64	&	+38 46 58.8	&	18.12	&	0.1	\\
Star 1	&		&	09 01 06.48	&	+38 47 08.7	&	18.14	&	1.4	\\
Star 2	&		&	09 01 05.15	&	+38 48 24.5	&	17.36	&	1.3	\\
J094533.99+100950.1	&	2011.2.10	&	09 45 33.99	&	+10 09 50.1	&	17.66	&	0.4	\\
Star 1	&		&	09 45 27.96	&	+10 08 47.8	&	16.89	&	0.4	\\
Star 2	&		&	09 45 37.93	&	+10 08 08.9	&	18.01	&	0.7	\\
J094533.99+100950.1	&	2012.1.28	&	09 45 33.99	&	+10 09 50.1	&	17.66	&	0.4	\\
Star 1	&		&	09 45 27.96	&	+10 08 47.8	&	16.89	&	0.4	\\
Star 2	&		&	09 45 37.93	&	+10 08 08.9	&	18.01	&	0.7	\\
J125219.48+264053.9	&	2011.4.23	&	12 52 19.48	&	+26 40 53.9	&	17.70	&	0.2	\\
Star 1	&		&	12 52 27.12	&	+26 38 49.7	&	17.44	&	1.1	\\
Star 2	&		&	12 52 14.26	&	+26 39 11.5	&	17.15	&	1.3	\\
J125219.48+264053.9	&	2013.4.14	&	12 52 19.48	&	+26 40 53.9	&	17.70	&	0.2	\\
Star 1	&		&	12 52 23.02	&	+26 38 42.9	&	15.82	&	0.6	\\
Star 2	&		&	12 52 23.82	&	+26 41 42.6	&	16.43	&	0.3	\\
J132809.59+545452.8	&	2011.4.24	&	13 28 09.59	&	+54 54 52.8	&	17.84	&	0.1	\\
Star 1	&		&	13 27 58.21	&	+54 54 00.2	&	17.54	&	0.8	\\
Star 2	&		&	13 28 22.83	&	+54 55 54.7	&	18.14	&	0.5	\\
J134601.29+585820.1	&	2011.4.25	&	13 46 01.29	&	+58 58 20.1	&	17.74	&	0.2	\\
Star 1	&		&	13 46 06.60	&	+58 58 08.2	&	18.01	&	1.5	\\
Star 2	&		&	13 45 55.76	&	+58 57 34.8	&	18.46	&	1.3	\\
J142943.64+385932.2	&	2013.4.13	&	14 29 43.64	&	+38 59 32.2	&	17.55	&	0.0	\\
Star 1	&		&	14 29 39.99	&	+39 02 19.6	&	17.21	&	0.9	\\
Star 2	&		&	14 29 30.47	&	+39 00 08.7	&	16.33	&	0.6	\\

\hline
\end{tabular}
\end{center}

\end{table*}

The photometric data were reduced with the standard routines in the
Image Reduction and Analysis Facility (IRAF) software. The bias
frames were extracted from no fewer than ten frames, and the flat frames
did not have fewer than five frames in one band for one night of observation.
The dark of the CCD is negligible (compared with the readout noise
and the flat fluctuation) and therefore not considered.
The flat frames for the same band were combined by average, and then the
normalized flat frame was generated; the normalized bias frame was
generated by median combination. Then the source frames
were corrected by the normalized bias frame and flat frame.

With the corrected source images, we used the package APPHOT
to perform aperture photometry. The values of \emph{enclosed},
\emph{moffat}, and \emph{direct} for the comparison stars and the
target source were used to estimate the mean full width at half
maximum (FWHM). The apertures of the photometry for individual
frame were carried with $2.5\sim3$ times of FWHM. If the value of FWHM significantly changed during one night, we took different values of
FWHM even for the same source.

\section{Results}
To detect the underlying variation of the target, we first
calculated the differential light curves (DLCs) between the target
and comparison stars. Two nearby comparison stars (noted as Star~1 and Star 2
hereafter) with similar magnitudes to the target were selected and
the DLCs of AGN$-$Star~1, AGN$-$Star~2, and Star~1$-$Star~2 are shown in
Figure 1.  The position and $g-r$ color of targets and comparison stars are shown in Table 2. Since the color difference between target and star pairs is smaller than 1.5, the variation in air mass during the observation has little effect on DLCs (Carini et al. 1992; Stalin et al. 2004). We also tried different companion stars, and the final significance of intranight variation is quite robust. Some exposures with bad weather were excluded from
the DLCs, which led to some gaps in the DLCs.

 \begin{figure*}[!htp]
 \centering
   \includegraphics[width=0.45\textwidth]{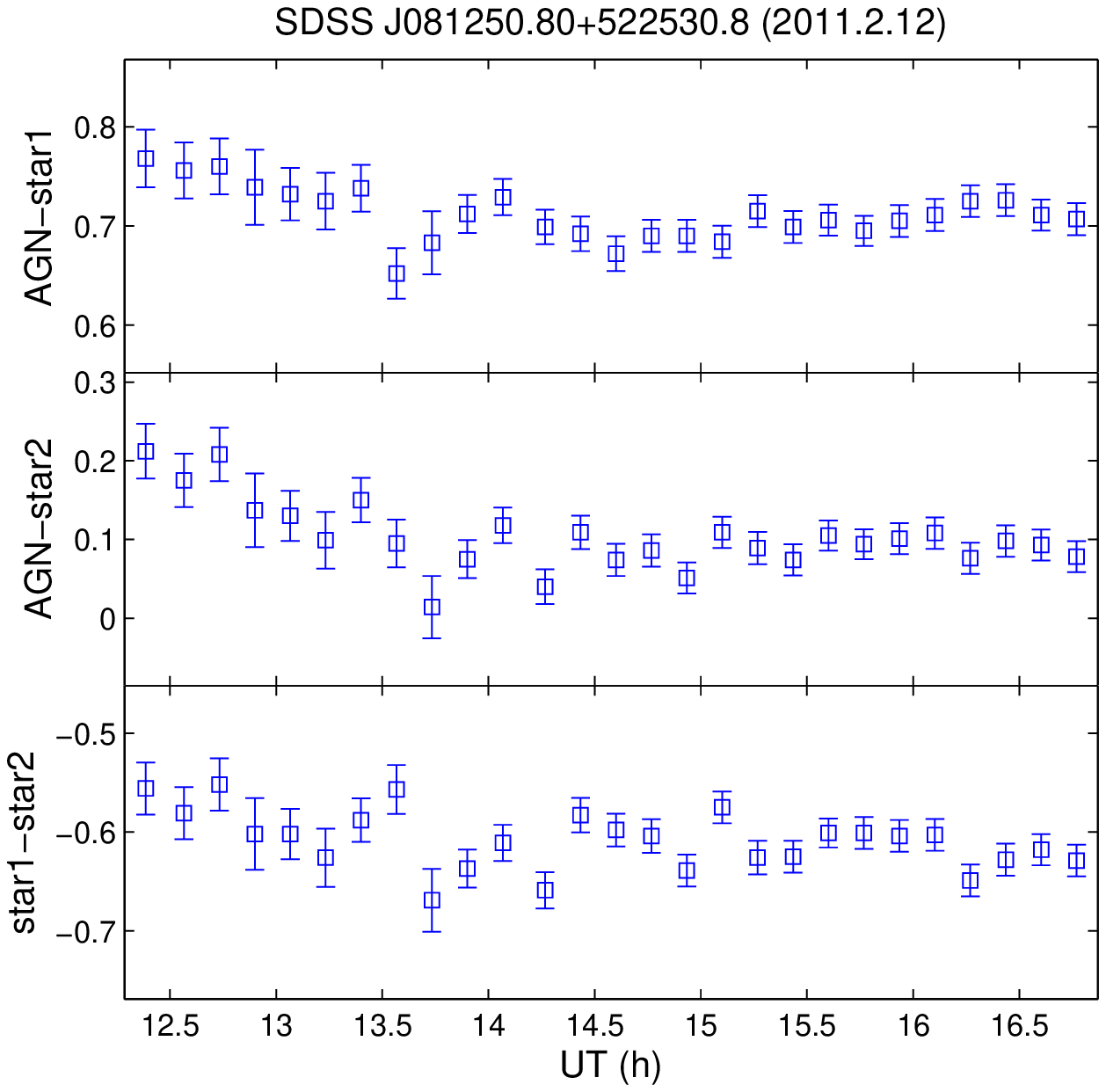}
   \includegraphics[width=0.45\textwidth]{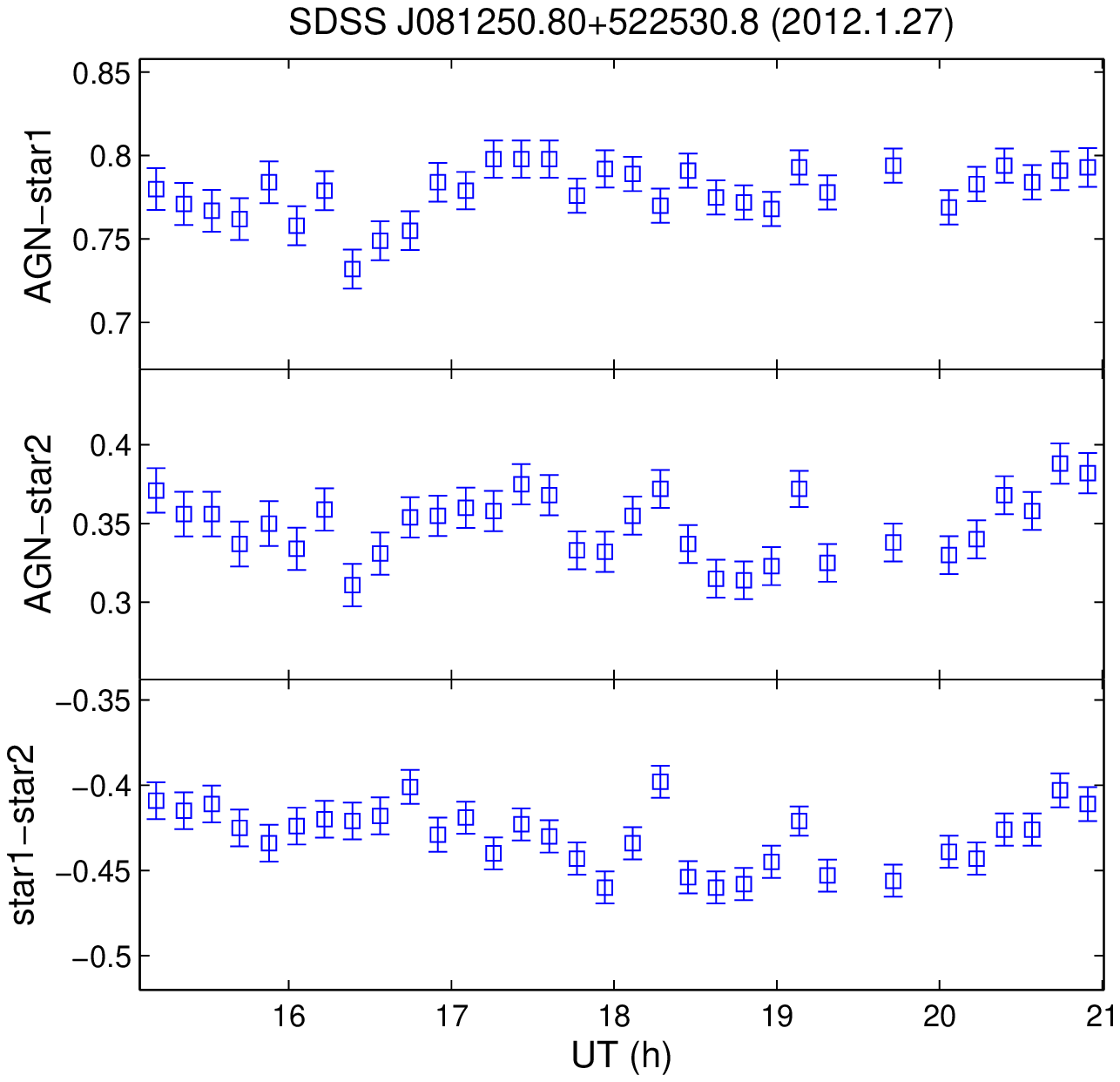}
   \includegraphics[width=0.45\textwidth]{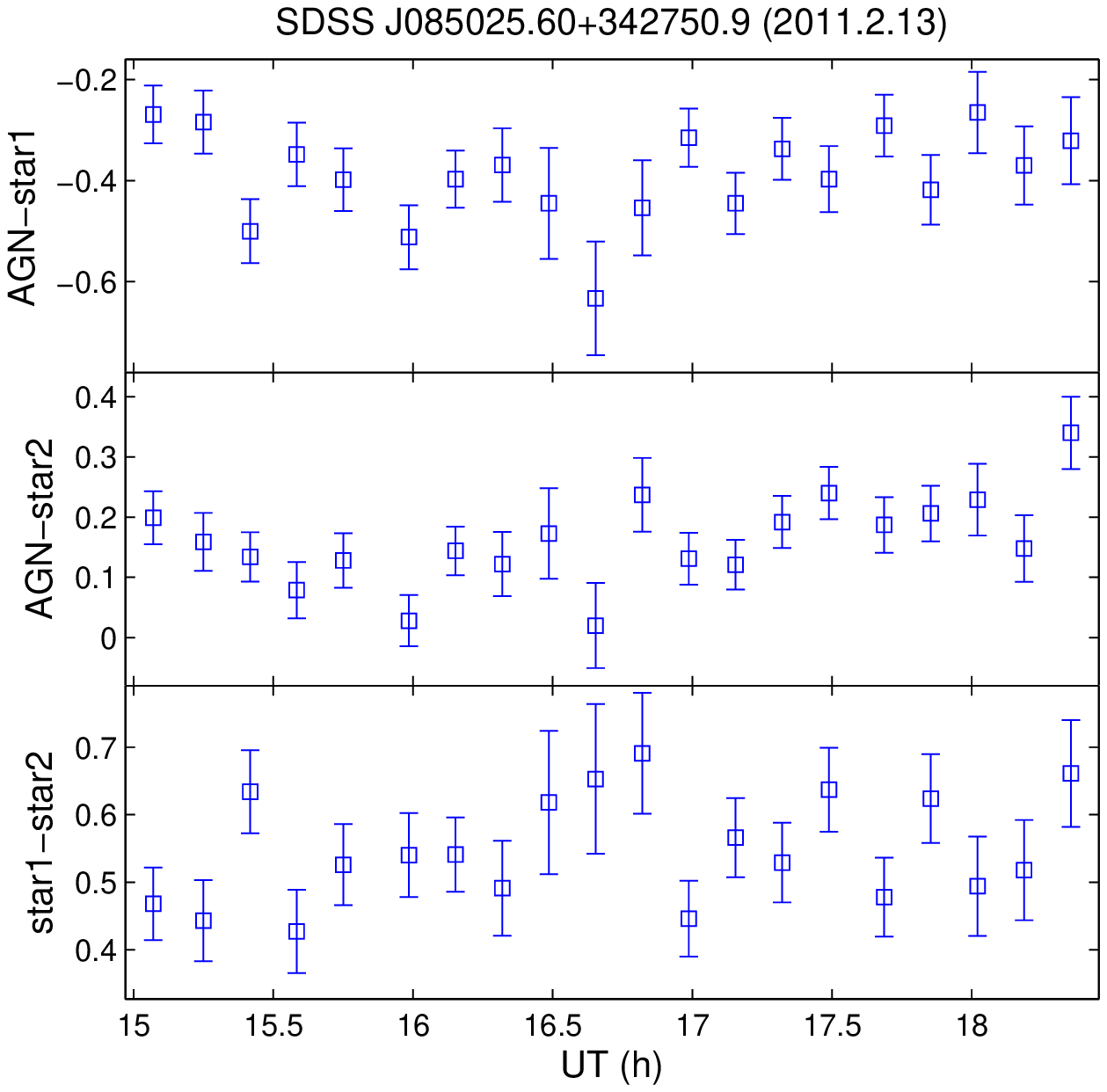}
   \includegraphics[width=0.45\textwidth]{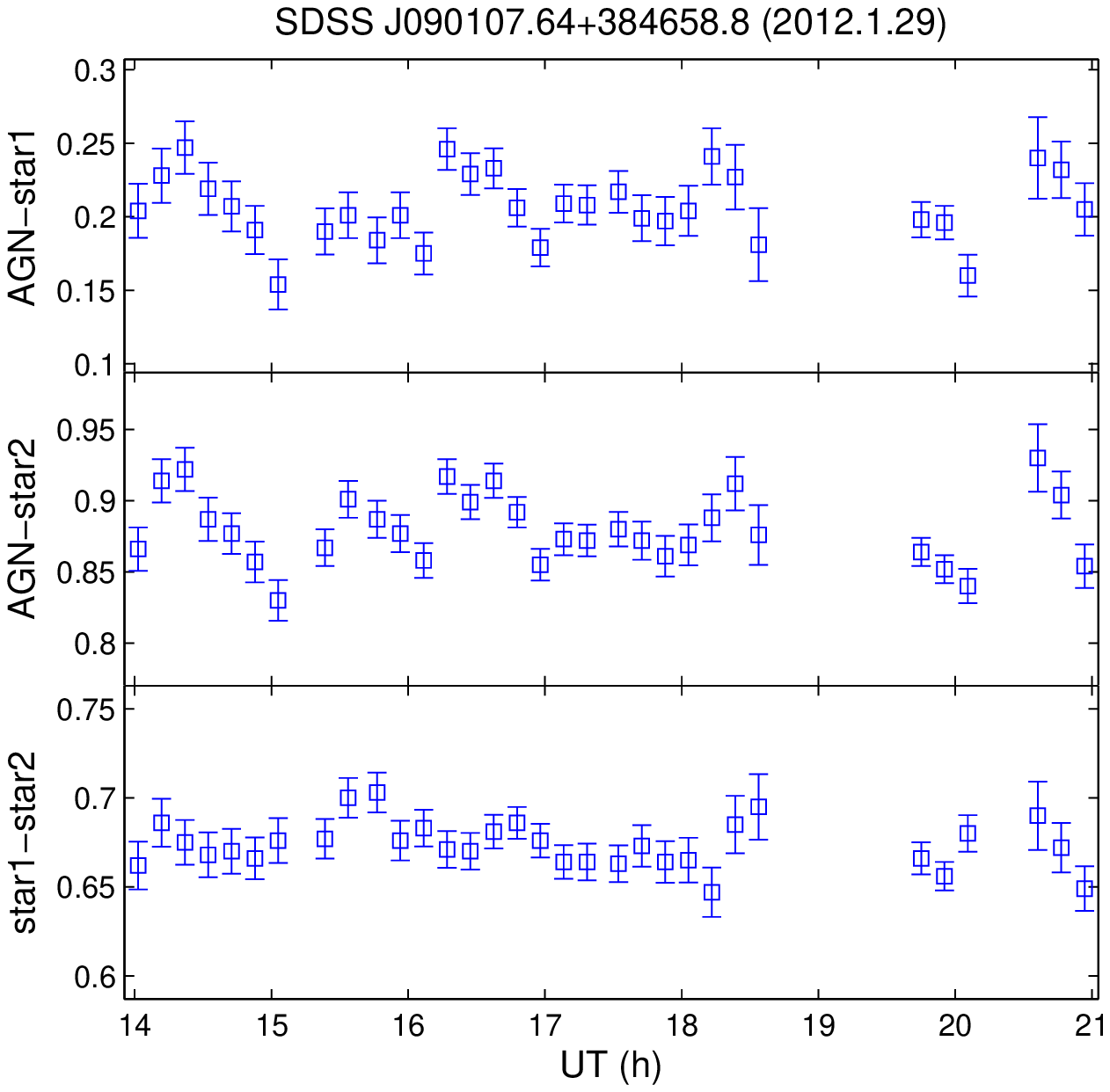}
   \includegraphics[width=0.45\textwidth]{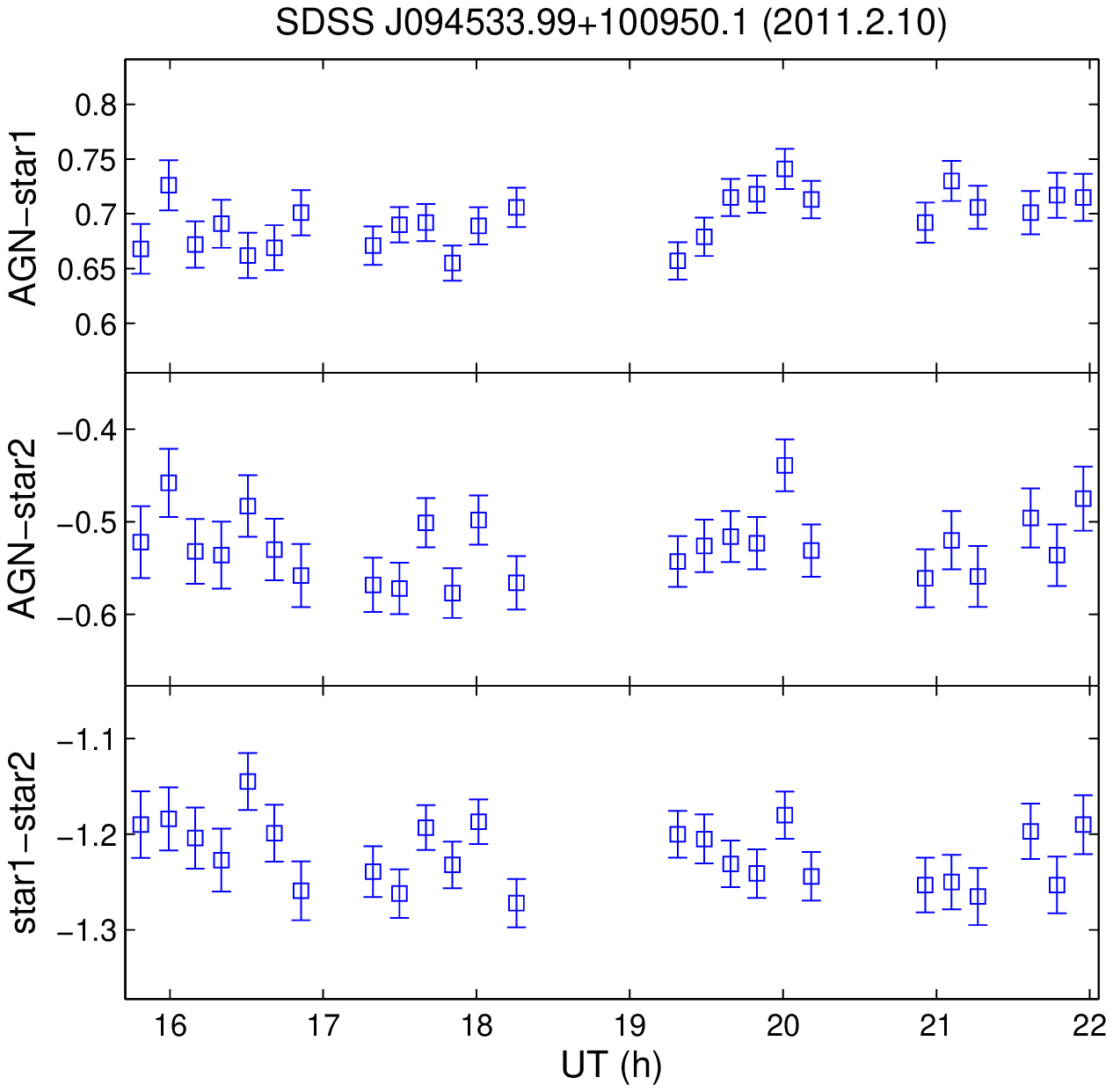}
   \includegraphics[width=0.45\textwidth]{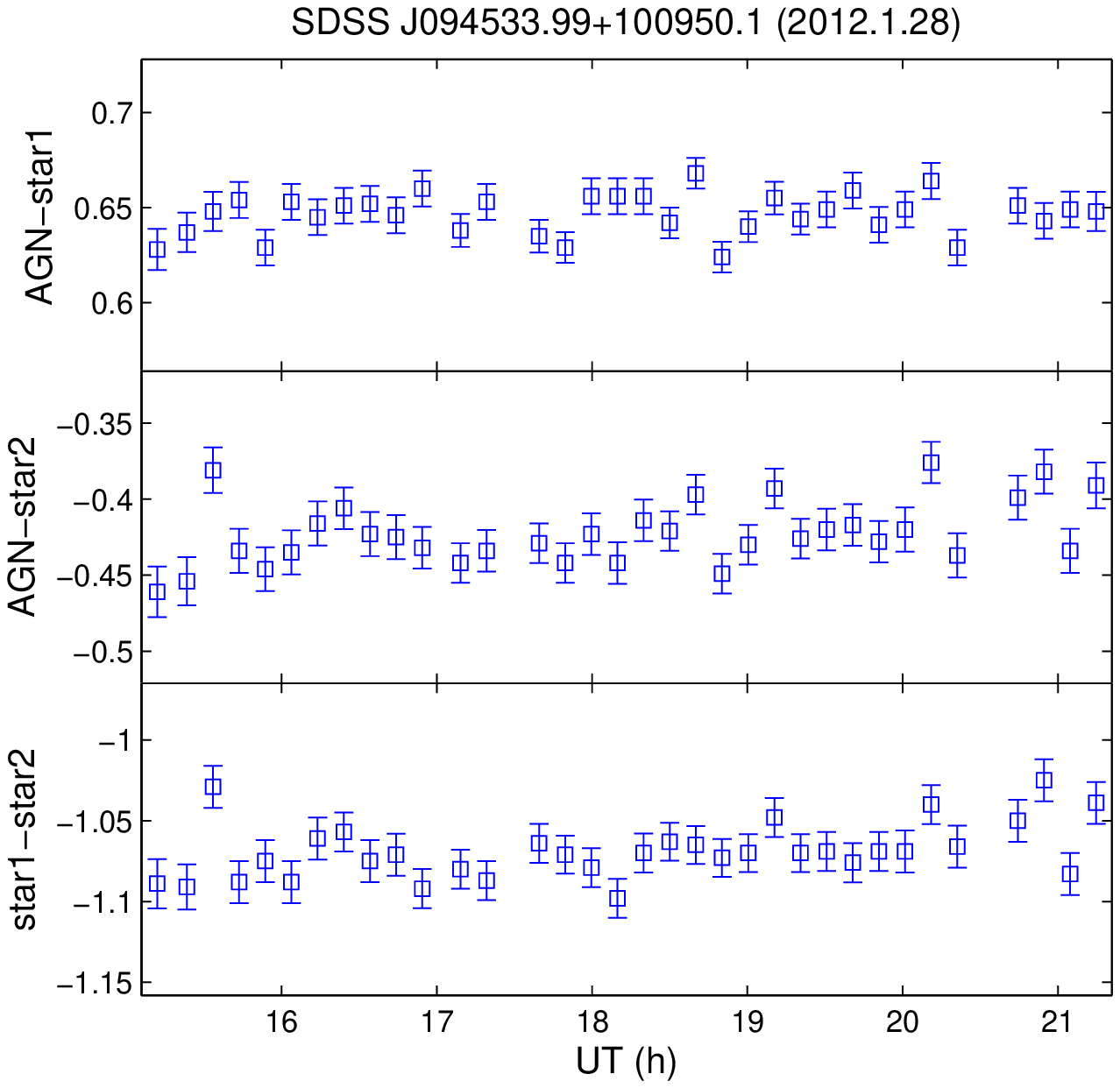}
   \caption{Differential light curves of AGN-Star~1, AGN-Star~2, and Star~1-Star~2.}
 \label{fig:1}
 \end{figure*}

 \begin{figure*}[!htp]
  \centering
   \includegraphics[width=0.45\textwidth]{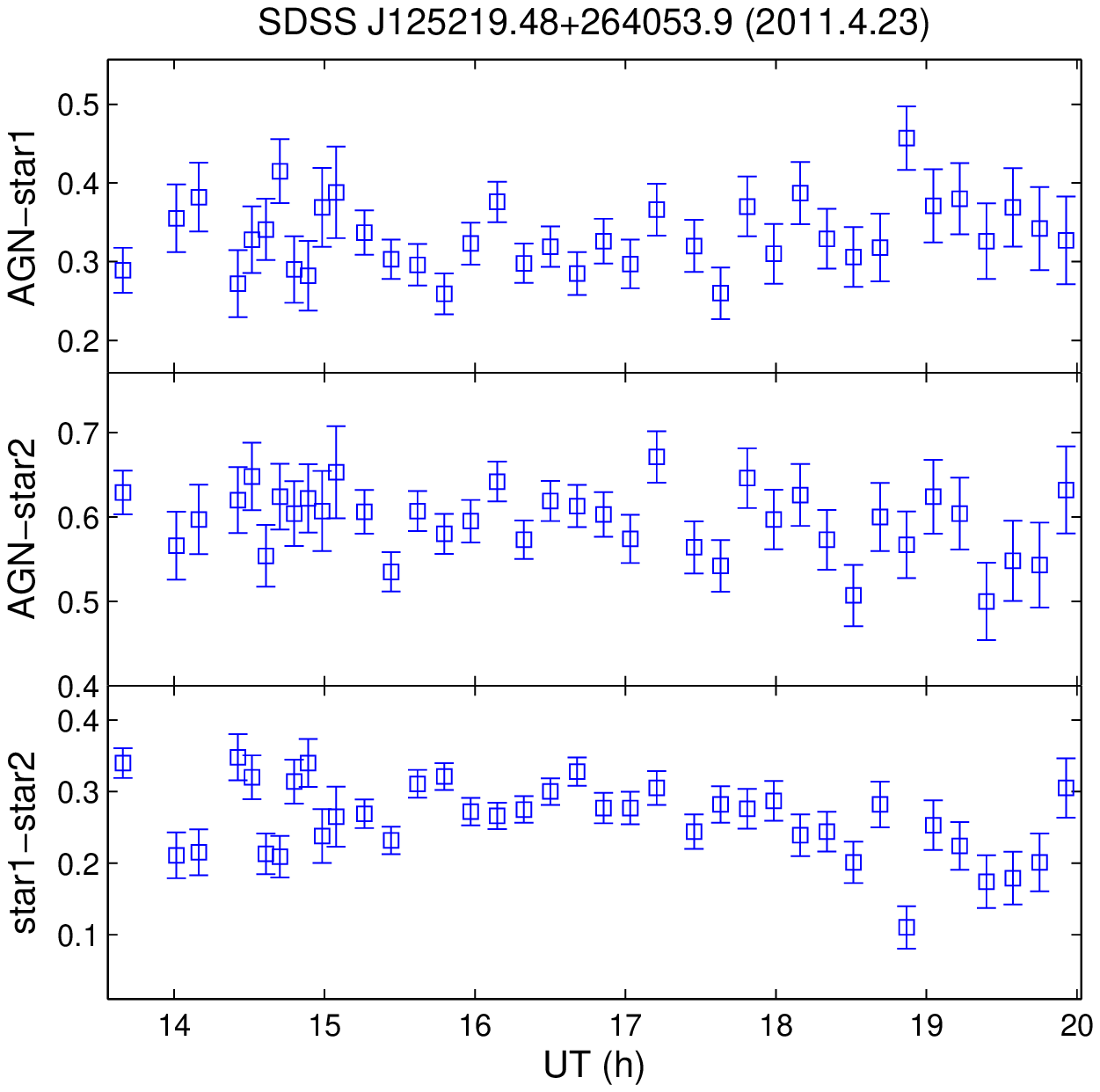}
   \includegraphics[width=0.45\textwidth]{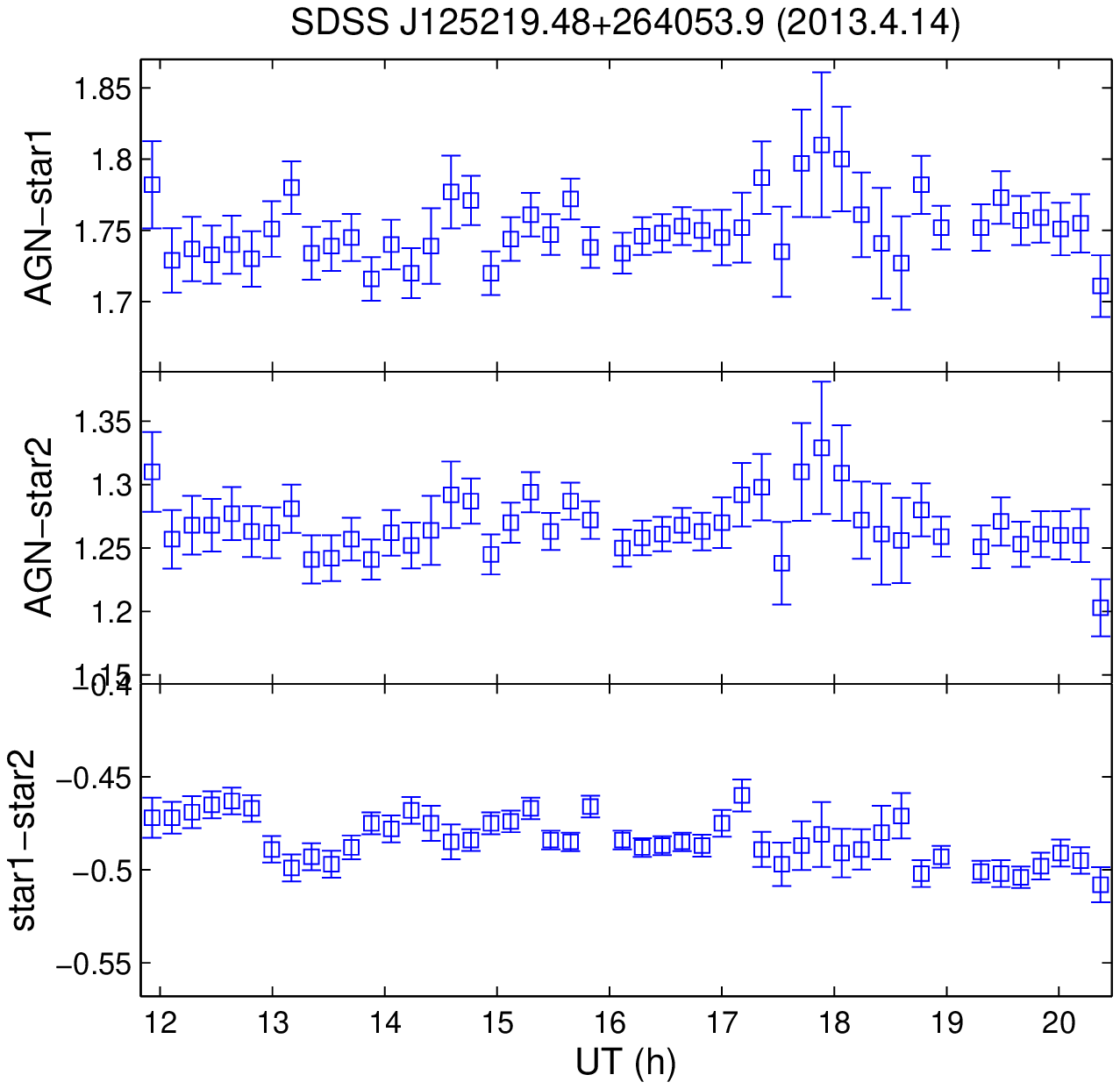}
   \includegraphics[width=0.45\textwidth]{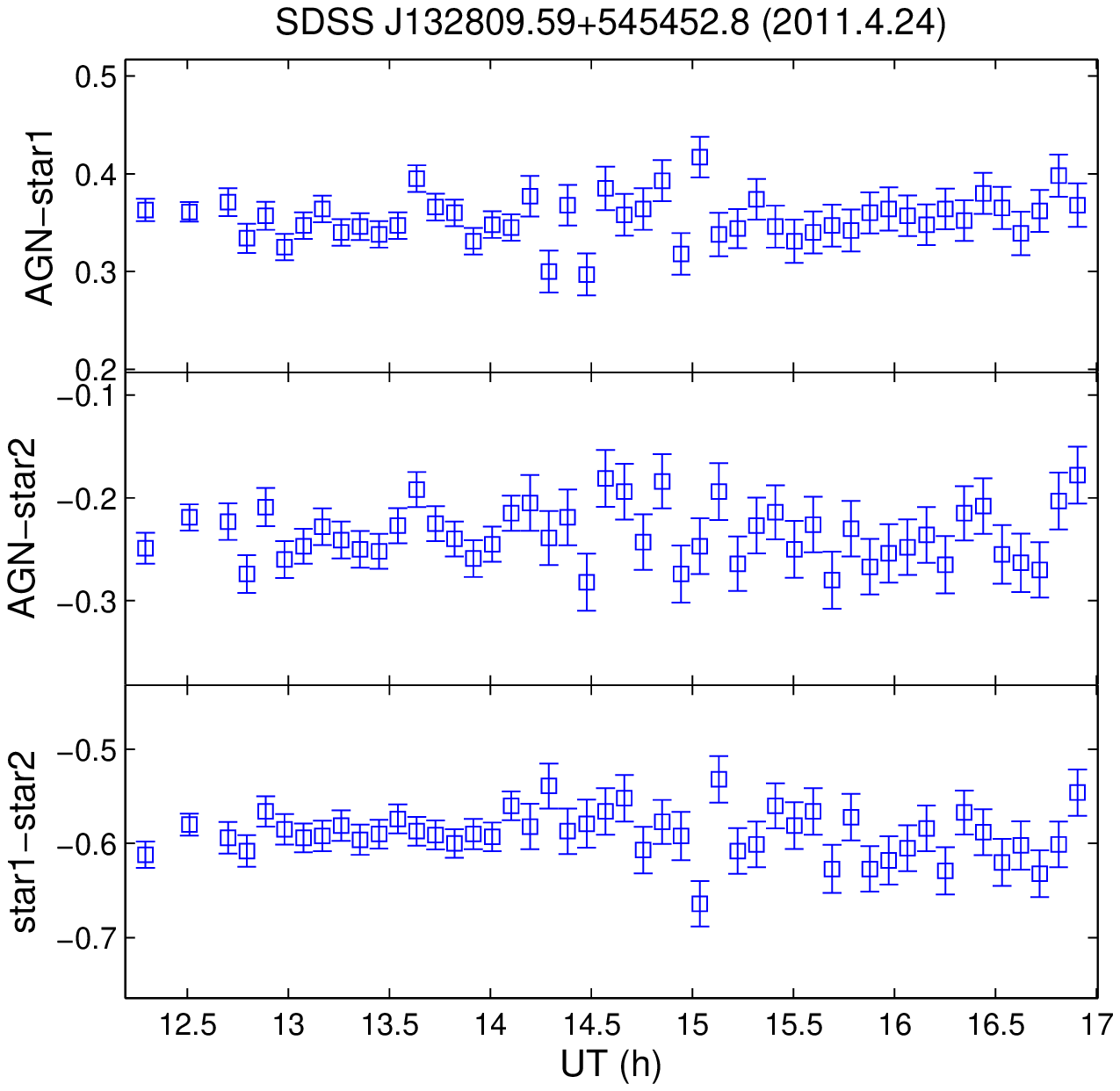}
   \includegraphics[width=0.45\textwidth]{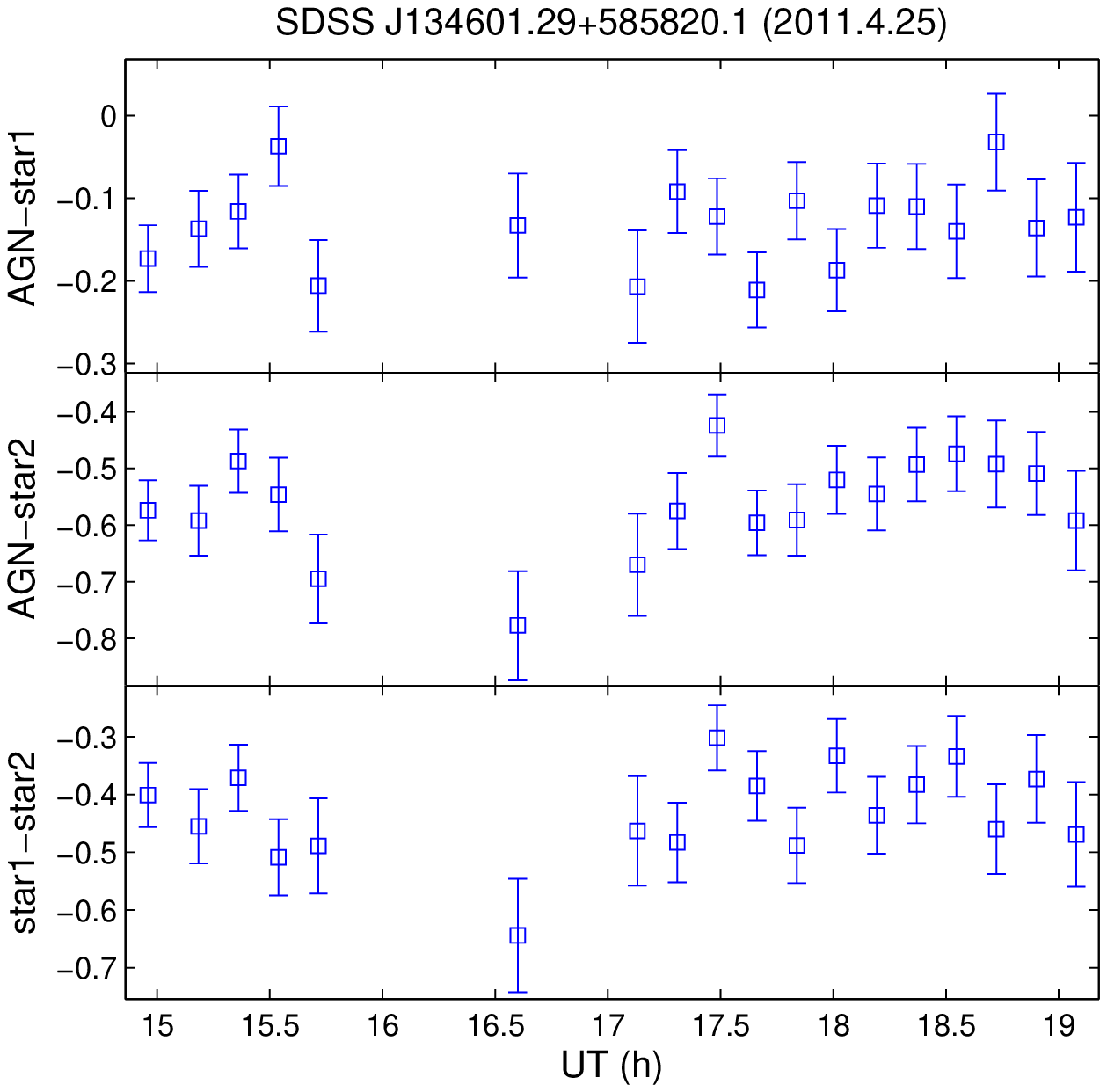}
   \includegraphics[width=0.45\textwidth]{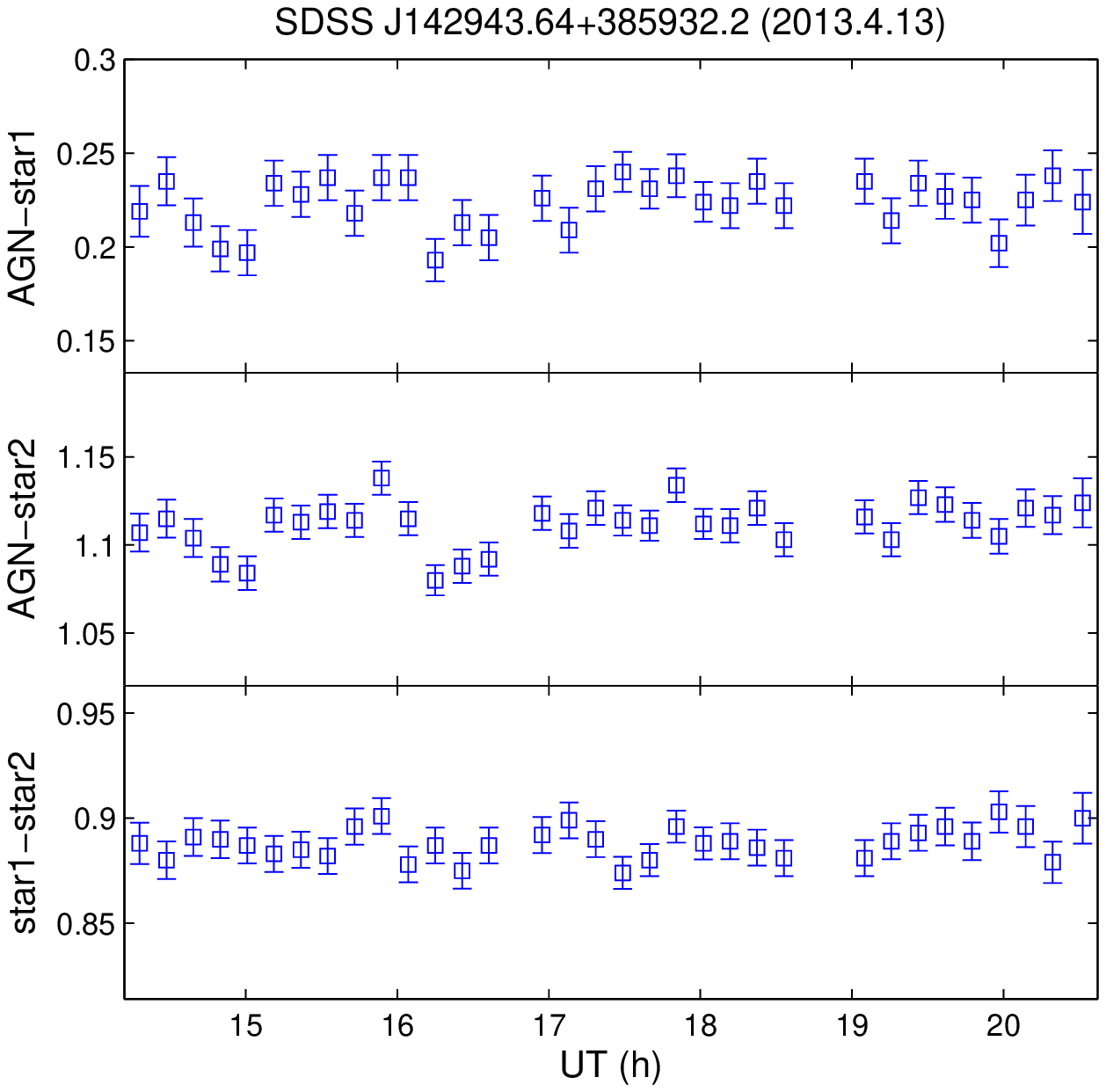}

    {{\bf Fig.~\ref{fig:1}.} \textit {continued}}

 \end{figure*}

To quantify the significance of the variation of light curves, we
performed a scaled $F$-test, which is more powerful and reliable
than the traditional $C$-test (de Diego 2010). The scaled $F$ value (Howell et al. 1988) is defined
as
   $$
     {F} = \frac{{s_{{\rm{AGN - Star~1}}}^2}}{{{\Gamma ^2}s_{{\rm{Star~1 - Star~2}}}^2}}, \eqno{(1)}
   $$
where $s_x^2 = \frac{1}{N-1}\sum\limits_{i = 1}^N {{{({X_i} - \bar
X)}^2}} $, and $x$ can stand for AGN-Star~1 or Star~1-Star~2.

The definition of $\Gamma ^2$ is
$$
{\Gamma ^2} = {\left(
{\frac{{{N_{{\rm{Star~2}}}}}}{{{N_{{\rm{AGN}}}}}}} \right)^2}\left[
{\frac{{N_{{\rm{Star~1}}}^2(N_{{\rm{AGN}}}^{} + P) +
N_{{\rm{AGN}}}^2(N_{{\rm{Star~1}}}^{} +
P)}}{{N_{{\rm{Star~2}}}^2(N_{{\rm{Star~1}}}^{} + P) +
N_{{\rm{Star~1}}}^2(N_{{\rm{Star~2}}}^{} + P)}}} \right], \eqno{(2)}
$$
which is the scaled factor to account for the different accuracies between
the photometries of the target and comparison stars (Howell et al.
1988). The variables $N_{\rm{AGN}}$, $N_{\rm{Star~1}}$, and $N_{\rm{Star~2}}$ are
the total counts (sky-subtracted) of target, Star~1, and Star~2,
respectively. The variable $P$ is defined as $P=n_p(N_S+N_r^2)$, where $n_p$ is
the number of pixels in the applied measuring aperture, the variable $N_S$ is the
sky photons per pixel, and $N_r$ is the readout noise ($e^-$/pixel).
The value of $\Gamma ^2$ can be calculated frame-by-frame. However,
the variation in $\Gamma ^2$ of our observations during one night is no
more than 10\% owing to the small variation of our targets.
Therefore, we have taken the median value of $\Gamma ^2$ for the exposures
in one night. Our final result is not sensitive to this choice.

The significance of the variation is determined by the $F$
distribution with $N_{\rm{AGN-Star~1}}-1$ and
$N_{\rm{Star~1-Star~2}}-1$ degrees of freedom, where
$N_{\rm{AGN-Star~1}}$ and $N_{\rm{Star~1-Star~2}}$ are the number of
observations in the AGN-Star~1 and Star~1-Star~2 DLCs, respectively.

%
%

The results of the significance are listed in Table 3. Since we exchanged
the position of Star~1 and Star~2 in equations (1) and (2), there are two
values of significance for ${\rm{(AGN - Star~1)/(Star~1 - Star~2)}}$
and ${\rm{(AGN - Star~2)/(Star~2 - Star~1)}}$.

As indicated by the results of  $F$-test, we only detect a significant variation ($\sim$3$\sigma$ level) in SDSS J090107.64+384658.8. However, due to the large proper motion of this source (62$\pm$11 mas/yr from Monet et al. 2003), its extragalactic nature is doubtable. Therefore, we would like to exclude it from the final sample of radio-quiet BL
Lac objects. As a result, there is no significant variation detected in our observations of radio-quiet BL Lac objects.

\begin{table*}[!t]
\caption{Results of the significance of variations.}
\label{table:1}
\begin{center}
\begin{tabular}{c c c c c c c c}
\hline\hline
Object (SDSS) &  Date    & $\Gamma_{1}$  &   $\Gamma_{2}$  &   $F_{1}$  &   $F_{2}$  &   Significance$_{1}$    &   Significance$_{2}$    \\
\hline
J081250.80+522530.8	&	2011.2.12	&	1.16	&	1.66	&	0.70	&	1.34	&	18.4\%	&	76.9\%	\\
J081250.80+522530.8	&	2012.1.27	&	1.50	&	1.85	&	0.51	&	0.72	&	3.44\%	&	18.7\%	\\
J085025.60+342750.9	&	2011.2.13	&	1.09	&	0.59	&	1.20	&	1.41	&	64.9\%	&	77.0\%	\\
J090107.64+384658.8	&	2012.1.29	&	1.84	&	1.34	&	1.75	&	2.68	&	93.8\%	&	99.6\%	\\
J094533.99+100950.1	&	2011.2.10	&	0.45	&	1.25	&	1.15	&	0.93	&	63.5\%	&	43.0\%	\\
J094533.99+100950.1	&	2012.1.28	&	0.51	&	1.26	&	0.74	&	1.16	&	20.1\%	&	66.2\%	\\
J125219.48+264053.9	&	2011.4.23	&	1.80	&	1.58	&	0.39	&	0.36	&	0.25\%	&	0.13\%	\\
J125219.48+264053.9	&	2013.4.14	&	6.76	&	7.15	&	0.49	&	0.46	&	0.91\%	&	0.50\%	\\
J132809.59+545452.8	&	2011.4.24	&	0.79	&	1.29	&	1.05	&	0.91	&	56.6\%	&	37.3\%	\\
J134601.29+585820.1	&	2011.4.25	&	0.60	&	0.96	&	0.67	&	1.18	&	20.9\%	&	63.1\%	\\
J142943.64+385932.2	&	2013.4.13	&	2.05	&	1.40	&	1.51	&	2.23	&	87.4\%	&	98.7\%	\\

\hline
\end{tabular}
\end{center}
\tablefoot{ The subscripts `1' and `2' of variables stand for the
results of ${\rm{(AGN - Star~1)/(Star~1 - Star~2)}}$} and ${\rm{(AGN -
Star~2)/(Star~2 - Star~1)}}$, respectively.

\end{table*}


\section{Discussions and conclusions}

Radio-loud AGNs and blazars can exhibit microvariation with a
large amplitude up to $\sim$100\%. However, some microvariation
events are also observed in radio-quiet AGNs with high significance
(Stalin et al. 2004; Gupta \& Joshi 2005). The mechanism of the microvariation in radio-loud AGNs is believed to be the fluctuation caused by the shocks in jets.
However, the instability or flares in the accretion disk can also
induce microvariation even for the radio-quiet AGNs (Mangalam
\& Wiita 1993). A weak
blazar component in radio-quiet AGNs is an alternative to
microvariation (Czerny et al. 2008). Though the occurrence of microvariations is not a smoking gun of jets, the fraction and
amplitude of the microvariations in radio-quiet and radio-loud ones
are quite different.

Gupta \& Joshi (2005) compiled the microvariations
of different classes of AGNs and found the detection fractions of
microvariation in radio-quiet and radio-loud (non-blazars) AGNs are
$\sim$10\% and $\sim$35-40\%, respectively. For blazars, the
fractions are $\sim$60-65\% and $\sim$80-85\% for the observations that are
less than and more than 6 h, respectively. In addition, they also
claim that the amplitude of the microvariation of radio-loud ones is
larger than that of radio-quiet ones.

Carini et al. (2007) established a sample of 117 radio-quiet AGNs
that have been investigated for microvariations and found a
detection rate of microvariations for the entire sample of 21.4\%.
If the criteria for `radio-quiet' are strengthened to $R<1$ ($R$ is
the ratio of the radio [5 GHz] flux to optical [4400 ${\AA}$] flux),
the detection rate of microvariations is only 15.9\%.

Goyal et al. (2013) analyzed 262 sessions of light curves of 77 AGNs from their uniform AGN monitoring data and found the duty cycles of intranight variation of radio-quiet quasars, radio-intermediate quasars,
lobe-dominated quasars, low optical-polarization core-dominated quasars, high optical-polarization core-dominated quasars, and TeV blazars are 10\%, 18\%, 5\%, 17\%, 43\%, and 45\%, respectively.

No significant microvariation is detected in our final sample, in ten sessions of light curves. The 1$\sigma$ upper limit of the fraction of microvariation is 15\%  using the method of Cameron (2011)\footnote{Given the sample size $n$ and observed success counts $k$, the upper limit $p_u$ is defined by $\int_{{p_u}}^1 {\frac{{(a + b - 1)!}}{{(a - 1)!(b - 1)!}}{p^{a - 1}}{q^{b - 1}}dp}  = (1 - c)/2$, where $a=k+1$, $b=n-k+1$,  $q=1-p$, and $c$ is the confidence level.}. In deriving this upper limit, we treated the sources equally. The weights of sources should not be the same owing to different exposure times, signal-to-noise ratios, and observation numbers; however, this potential minor correction will not change our final conclusion. This low fraction in our sample of radio-quiet BL
Lac objects is consistent with that of the radio-quiet AGNs but much
lower than for the radio-loud ones and blazars. This indicates
that the continuum of radio-quiet BL Lac objects is not dominated by
 the jet component. Actually, the SED of radio-quiet BL Lac objects is similar to
  the normal radio-quiet AGNs (Lane et al. 2011), which further supports the accretion disk origin of the continuum. Accurate black hole mass measurements can determine the accretion state of radio-quiet BL Lac objects and further distinguish the different models related to the accretion disk origin of their continua,  which will be explored in our future works.

GJC2013 and CKG2014 detected significant variations (confidence level $>$99\% for two comparison stars) of SDSS J090843.25+285229.8 and SDSS J121929.45+471522.8 in their 29 light curves. Based on these two events, they derived a duty cycle $\sim$5\% of intranight variation from their sample on weak-line AGNs. However, two sources in their sample (SDSS J090107.64+384658.8 and SDSS J121929.45+471522.8) are likely to be galactic sources owing to the large proper motion. The proper motions of SDSS J090107.64+384658.8 and SDSS J121929.45+471522.8 from USNO-B are 62$\pm$11 mas/yr and 112$\pm$4 mas/yr, respectively (Monet et al. 2003). These two bright sources ($V$$\sim$18) are well above the completeness limit $V$=21 of the USNO-B catalog (Monet et al. 2003). The positional error of one epoch is $\sim$200 mas. Thus, the proper motion of $\sim$100 mas/yr can be accurately detected with the epoch difference of $\sim$40 years. The systematic error of the proper motion is comparable to the statistical error. Munn et al. (2004) and Roeser et al. (2010) have further calibrated the USNO-B catalog with SDSS and 2MASS astrometry, and the resulting proper motions of these two sources are consistent with those from the USNO-B catalog. No radio or X-ray counterpart is found near their positions. Actually, Rebassa-Mansergas et al. (2010) lists SDSS J121929.45+471522.8 as a DC white dwarf in their catalog. SDSS J090107.64+384658.8 is classified as an uncertain DC white dwarf by Eisenstein et al. (2006) and as an uncertain DC+M binary system by Kleinman et al. (2013). The variation observed is likely due to the oscillation or accretion of the white dwarfs (Winget \& Kepler 2008; Fontaine \& Brassard 2008).

The DC white dwarfs are generally difficult to positively identify due to their featureless spectra and the discrepancy indeed exists in different catalogs. However, we think these possible galactic sources should be excluded from the sample of AGNs to be safe. We detected the intranight variation in SDSS J090107.64+384658.8 but excluded it from the sample.
 After the possible white dwarfs are removed from the sample of GJC2013 and CKG2014, only one significant variation (SDSS J090843.25+285229.8) is detected in the remaining 22 light curves, which is also the only significant event in all 32 light curves (including 10 additional ones of this paper) of weak-line AGNs up to now. This low occurrence rate is consistent with the rate for the normal radio-quiet AGNs.

Our present sample of radio-quiet BL Lac objects is still too small
to constrain the duty cycle of the microvariation well. We will
enlarge our sample, especially for the monitoring time
longer than six hours, and improve the accuracy of photometry to $\sim$
0.01 mag to detect smaller variations.

Observations in more bands will help to investigate the
color-behavior and further constrain the mechanism of the continuum
of the radio-quiet BL Lac objects.

\begin{acknowledgements}
The authors thank the referee for useful comments that 
improved the paper. This work is supported by 973 Program of China under grant
2014CB845802, by the National Natural Science Foundation of
China under grant Nos. 11103019, 11133002, and 11103022, and 11373036, and by the Strategic Priority Research Program ``The Emergence of Cosmological Structures" of the Chinese Academy of Sciences, Grant No. XDB09000000. We acknowledge the support of the staff of the Xinglong 2.16m and the Lijiang 2.4m telescope. Funding for the Lijiang 2.4m telescope has been provided by CAS and the People's Government of Yunnan Province. This work was partially supported by the Open Project Program of the Key Laboratory of Optical Astronomy, NAOC, CAS.
\end{acknowledgements}

\end{document}